# Incorporation of lanthanide ions into lead titanate

A. Peláiz-Barranco[1,*], Y. Méndez-González[1], D. C. Arnold[2], D. J. Keeble[3], P. Saint-Grégoire[4].

[1]Facultad de Física-Instituto de Ciencia y Tecnología de Materiales, Universidad de La Habana. San Lázaro y L, Vedado. La Habana 10400, Cuba.
[2]School of Chemistry, University of St Andrews, St Andrews, Fife KY16 9ST, United Kingdom.
[3]Carnegie Laboratory of Physics, School of Engineering, Physics, and Mathematics, University of Dundee, Dundee DD1 4HN, United Kingdom.
[4]University of Nîmes, Department of Sciences and Arts, 30021 Nimes cedex 01, and ICGM (UMR CNRS n° 5253), C2M, 34095 Montpellier cedex 05, France.

**Abstract**

A series of lanthanide, Ln, ion doped lead titanate ceramics, at 2 and 8 at.%, have been prepared with starting compositions $Pb_{1-3x/2}Ln_xTiO_3$, Ln = La, Nd, Sm, Eu, Gd, Dy, using a standard ceramic method. Structural, and differential scanning calorimetry, measurements are reported. Doping with $La^{3+}$ results in a reduction in tetragonality and cell volume; a partial recovery was then observed for the smaller $Ln^{3+}$ ions and suggests the onset of partial substitution of the rare earth on the B-site. The ferroelectric-paraelectric transition temperature results also consistent with an increased probability the B-site occupation with decreasing ion size.

**Keywords**: lanthanides, phase transition, perovskites.

## 1. Introduction

Pure lead titanate ($PbTiO_3$) would be a highly useful bulk piezoelectric material, due to its high intrinsic polarisation and relatively high Curie temperature, however, the large tetragonality prevents the processing of high density ceramic samples [1]. Lead titanate (PT) has the perovskite, $ABO_3$, structure with the $Pb^{2+}$ ions at the cell corner A-sites, $Ti^{4+}$ ions occupying the body centre B-sites, and the oxygen atoms at the face centre sites. It has been found that the PT lattice parameters can be reduced, while near optimal piezoelectric properties are retained, by appropriate doping. The inherent piezoelectric anisotropy depends on the degree of microstress or structural defects, including the pseudorandom distribution of oxygen defects, which are present in these materials before the poling process [2].

The perovskite structure has an intrinsic capability to host ions of different size, and a wide variety of elements can be accommodated. Interest has developed in the doping of perovskite oxide titanates with lanthanide ions (Ln). The Ln ion valence is normally fixed, but it has been proposed that the site of incorporation may vary with ion size, cation ratio, and oxygen environment during firing. Several studies have been performed on Ln incorporation to A- and/or B-sites of the perovskite structure in barium titanate ($BaTiO_3$) [3-6]. It has been reported that $La^{3+}$ and $Pr^{3+}$ ions prefer to substitute at the Ba site, whereas $Gd^{3+}$ and $Tb^{3+}$ partially substitute at the Ti site; for $Er^{3+}$ (and $Yb^{3+}$) preferential substitution at the Ti site has been predicted [6]. There is evidence lanthanide doping of $BaTiO_3$ improves resistance to electrochemical and aging failure [7].

---
[*] Corresponding Author: pelaiz@fisica.uh.cu

A previous study of $Ln^{3+}$ ion doped PT, co-doped with 2 at.% Mn, has provided experimental evidence consistent with the partial substitution of the smaller Ln ions $Dy^{3+}$, $Ho^{3+}$ and $Er^{3+}$ at the Ti site [8]. Two ferroelectric-paraelectric (FE-PE) phase transitions were inferred and analyzed by considering two different contributions to the macroscopic response; one due to Ln ions occupy the A-sites, and the other due to B-site occupancy [8]. However, for the larger lanthanide ions, $La^{3+}$, $Nd^{3+}$, $Sm^{3+}$ and $Gd^{3+}$, the macroscopic dielectric behaviour was attributed to a single FE-PE phase transition and the hopping of oxygen vacancies [9]. A theoretical analysis suggested that incorporation to B-site may be possible; however, the dielectric response did not provide evidence for partial Ti-site substitution for these ions [9].

This work details a study of lanthanide ion doping of pure lead titanate ceramics (without co-doping). Structural analysis and differential scanning calorimetry measurements were performed on 2 and 8 at% Ln doped samples, using ions varying in size from $Dy^{3+}$ to $La^{3+}$, to provide insight on the possible site of incorporation in the perovskite structure, and to study the effect on the ferroelectric-paraelectric transtion temperature.

## 2. Experiment

Dense ceramic samples were prepared by the standard solid state method using a nominal composition, that assumed donor doping, $(Pb_{1-3x/2}Ln_x)TiO_3$ with $x = 0.02$ and $0.08$, and Ln = La, Nd, Sm, Eu, Gd, Dy. The mixture of powders was prefired at 900 °C in air for 2 h, and sintering was carried out at 1220 °C for 2 h in a covered platinum crucible. The samples are labelled PTLn$x$, for example PTLa8 represents the 8 at.% La doped sample. X-ray diffraction (XRD) patterns, taken at room temperature, were collected in transmission mode using a STOE diffractometer and Cu radiation (40 kV and 40 mA, wavelength = 1.5418 Å). The differential scanning calorimetry (DSC) measurements were performed between 100 and 550 °C in argon atmosphere using a Netzsch DSC 404C and a TASC 414/4 controller.

## 3. Results and Discussion

From the nominal compositions used in the study it is expected that the Ln ions will substitute at the A-site, since all the $Ln^{3+}$ ions used have a crystal radii in 12 coordination smaller than $Pb^{2+}$ (163 pm) but when in 6 coordination are significantly larger than $Ti^{4+}$ at the B-site (74.5 pm) [10, 11]. In consequence, a decrease in tetragonality is expected on Ln ion doping. It should also be noted that charge balance requires that the local positive charge resulting from the substitution be compensated by the formation of cation vacancies, assumed to be Pb vacancies. If this is the case is then nominal composition is $(Pb_{1-3x/2}Ln_xV_{x/2})TiO_3$, with $x = 0.02$ and $0.08$, and $V$ representing a vacancy. However, should partial substitution for $Ti^{4+}$ occur then the tetragonality would tend to increase with increasing occupancy of a larger ion on the B-site. If the possibility is considered then composition takes the form $(Pb_{1-3x'/2}Ln_{x'}V_{x'/2})(Ti_{1-y}Ln_y)(O_{3-y/2}V_{y/2})$, with $x' + y = x$. Using this expression, it is possible to perform tolerance factor, $t$, calculations for the doped structures. This geometric factor analysis provides some insight on the effect of the Ln doping, here calculations were performed for $x = 0.08$ and varying the fraction, $y$, of this doping incorporated at the B-site, see Figure 1, all the values shown fall in the range for stable perovskite structures [12]. These tolerance factor calculations do not preclude the possibility of partial B-site substitution.

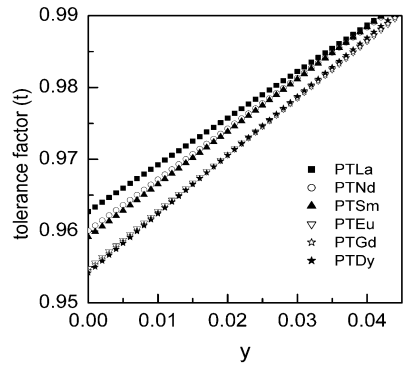

**Figure 1**. Calculated tolerance factor values assuming compositions $(Pb_{1-3x'/2}Ln_{x'}V_{x'/2})(Ti_{1-y}Ln_y)(O_{3-y/2}V_{y/2})$ with $x' + y = 0.08$, for varying partial B-site occupancy ($y$) for the lanthanide ions studied.

The XRD patterns for the 2 and 8 at.% doped samples, as a function of Ln ion dopant ordered by decreasing ionic radius, are shown in figure 2 and figure 3, respectively. In all cases the main phase can be indexed to $PbTiO_3$ in the space group P4mm. The diffraction patterns for PTSm2 and PTGd8 show small peaks from a secondary phase (marked by using arrows), which could be identified as $SmTi_2O_7$ and $Gd_2Ti_2O_7$, respectively. The peaks marked with an asterisk are due to the XRD collimator.

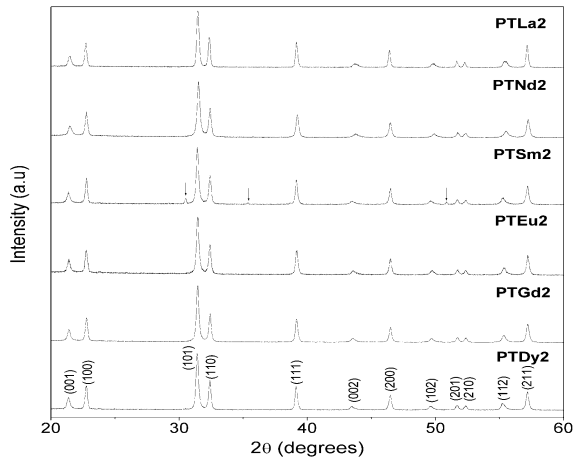

**Figure 2.** X-ray diffraction patterns for the 2 at.% lanthanide doped lead titanate samples. The arrows for PTSm2 mark peaks assigned to the impurity phase $Sm_2Ti_2O_7$.

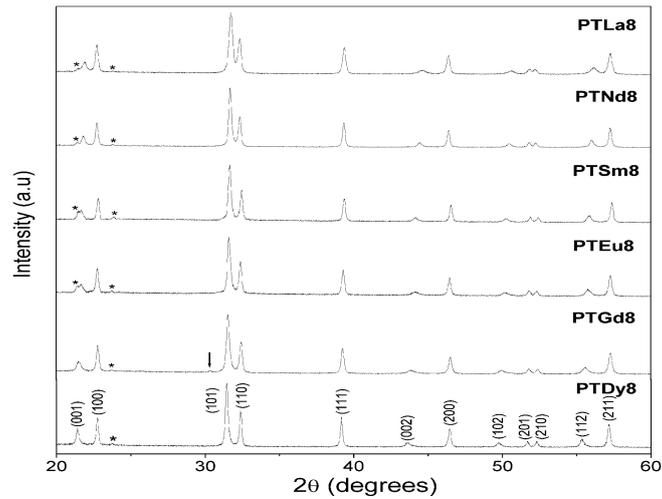

**Figure 3.** X-ray diffraction patterns for the 8 at.% lanthanide doped lead titanate samples. The arrow for PTGd8 mark peak assigned to the impurity phase $Gd_2Ti_2O_7$. Peaks marked with an asterisk are due to the XRD collimator.

The lattice parameters, tetragonality, and cell volume, for the 2 and 8 at.% doped samples are shown in figure 4 and figure 5, respectively; values for pure $PbTiO_3$, obtained under the same conditions are also shown. All these quantities decrease in value between $PbTiO_3$ and the La-doped samples. For example, the cell volume reduces from 63.5 Å$^3$ in PT to 63.1 and 62.6 Å$^3$ for PTLa2 and PTLa8, respectively; similarly the tetragonality decreases from 1.063 to 1.058 and 1.029, respectively. These changes are consistent with a mechanism of substitution of $Pb^{2+}$ (163 pm) by the the smaller $La^{3+}$ ion (150 pm) at the A-site [10, 11].

Figure 4 shows that for the 2 at% doped samples the *a*–parameter decreases with decreasing ionic radius of the Ln ion, however, a small anomaly is observed for Sm. As noted above, XRD detects the presence of a secondary phase in this sample. The *c*–parameter initially decreases with decreasing dopant ion size down to Nd and then increases, recovering to the undoped value for Dy. Again, the Sm value is anomalous. These results are reflected in the *c*/*a* ratio and unit cell volume behaviours shown, the values decrease to Nd and then recover toward those for pure PT as the ion size reduces to Dy. In the case of the *c*/*a* ratio the PTDy2 sample value slightly exceeds that for PT, while the cell volume value is slightly less than that for the pure material.

The crystal radius of $Gd^{3+}$ and $Dy^{3+}$ in 12-coordination have not been reported, but will be less than 138 pm (the $Sm^{3+}$ value), and compared to the $Pb^{2+}$ radius of 163 pm. The 6-coordination radius values for $Gd^{3+}$ and $Dy^{3+}$ are 107.8 pm and 105.2 pm, respectively, which are still significantly larger than the $Ti^{4+}$ 6-coordination value of 74.5 pm [10, 11]. The recovery of the values shown in figure 4 with decreasing Ln ion crystal radius toward the those for pure PT, however, suggest $Gd^{3+}$ and $Dy^{3+}$ have a finite probability of substituting at the B-site. The recovery in tetragonality could be due to the recovery in the fraction of $Pb^{2+}$ ions at the A-site, but may also contain a contribution from increased distortion of the Ln ion substituted B-sites.

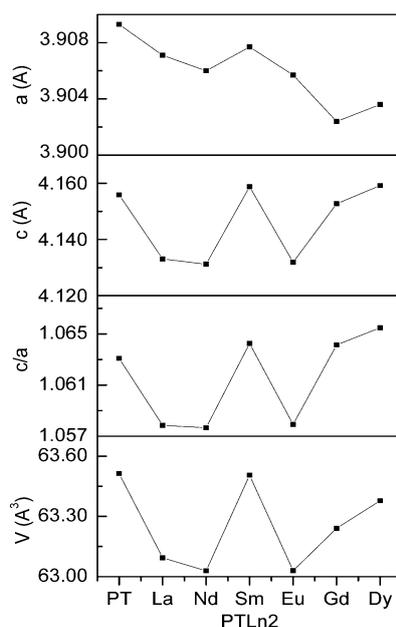

**Figure 4.** Lattice parameters (*a*,*c*), tetragonality (*c*/*a*), and unit cell volume for undoped $PbTiO_3$ and for the 2 at.% lanthanide doped lead titanate samples.

The lattice parameters, tetragonality, and lattice volume, for the 8 at.% doped samples shown in figure 5, apart from the *a*-parameter, exhibit markedly larger changes compared to the 2 at.% samples. The *a*-parameter increases between the undoped (UD) and the PTLa8 sample; it then decreases toward the UD value for Nd and then follows a similar trend to that seen for the 2 at.% samples. The *c*-parameter, however, shows a large (~ 3 %) decrease between the UD and PTLa8 sample, the value then begins to recover with decreasing dopant ion size. This behaviour of the *c*-parameter is directly reflected in that for the *c*/*a* ratio. The trend in the unit cell volume is similar to that seen for the 2 at.% doped sample, with the minimum occurring for Nd, and the decrease with respect to the UD value is ~0.8 % for PTNd2 and ~1.8 % for PTNd8, respectively.

Doping with La, at either 2 or 8 at.%, results in a marked contraction of the *c*-parameter and hence of the tetragonality and cell volume, which is consistent with the substitution of $La^{3+}$ ions at $Pb^{2+}$ ion sites. The *c*-axis contraction is comparable for Nd doping and, for both the 2 and 8 at.% samples, the unit cell volume reaches a minimum for this ion. The crystal radius at the A-site is reduced from 150 pm for $La^{3+}$, ~ 8 % smaller than the $Pb^{2+}$ crystal radius, to 141 pm for $Nd^{3+}$, ~14 % smaller. The 6 coordination radius for $Nd^{3+}$ is 112.3 pm, ~ 51 % larger than $Ti^{4+}$ [10, 11]. From figure 4 and figure 5 there is evidence that on reducing the ion size further, to Sm then to Eu, that the cell contraction is halted and, for the higher doping level, a slight recovery of the tetragonality value occurs. The rate of decrease of the 12 coordination crystal radius reduces; $Sm^{3+}$ has a radius of 138 pm, and that for $Eu^{3+}$ can be estimated to be ~ 135 pm [10, 11]. However, the trends in the structural data suggests that there is a finite probability of a fraction of $Eu^{3+}$ ions incorporate at the B site, despite the 6 coordination crystal radius being ~ 46 % larger than the $Ti^{4+}$ value. However, in considering the possibility of an increasing fraction of B-site substitution with decreasing ion size it should be noted that segregation to a second phase, such as $SmTi_2O_7$ observed for PTSm2 and $Gd_2Ti_2O_7$ observed for PTGd8, effectively removes dopant ions, and so could also contribute to a recovery of tetragonality seen with the smaller ions.

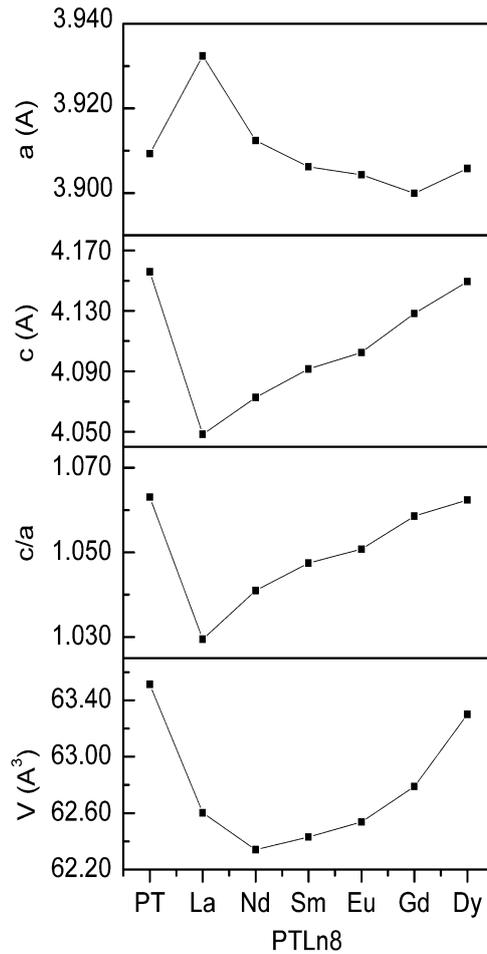

**Figure 5.** Lattice parameters (*a*,*c*), tetragonality (*c*/*a*), and unit cell volume for undoped $PbTiO_3$ and for the 8 at.% lanthanide doped lead titanate samples.

Figure 6 shows the differential scanning calorimetry results for PTLn2 samples. A clearly defined peak is normally observed. However, the PTLa2, PTNd2, and PTGd2, samples exhibit an additional thermal anomaly in the region of 440 °C; in the case of PTGd2 simply leads to a slight broadening of the peak on the low temperature side. The broadening and asymmetry around this temperature may be due to decrease in crystallinity, in this range, for these samples. The well defined peak corresponds to the ferroelectric to paraelectric phase transition temperature ($T_c$). This is observed to decrease from the pure PT value of 490 °C to 470 °C for PTLa2, then to continue to reduce until reaching a value of 445°C for PTEu2. This reduction in the FE-PE transition temperature is consistent with the substitution of the smaller $Ln^{3+}$ ions for $Pb^{2+}$ at the A-site. However, for the Gd and for Dy there is a recovery of $T_c$ to 460 °C and 465 °C, respectively, concomitant with the recovery in the values of the structural parameters for these samples toward those for UD PT (figure 4).

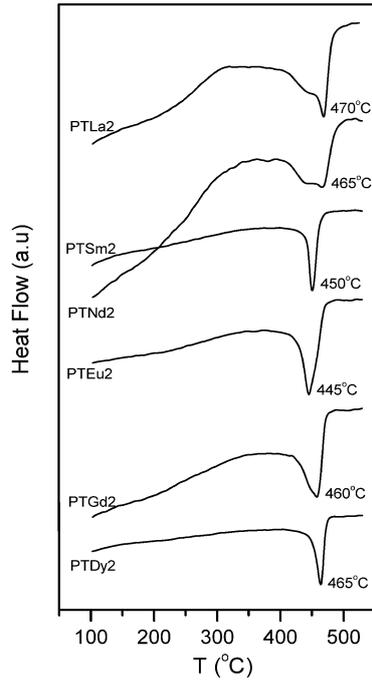

**Figure 6.** The differential scanning calorimetry curves for the 2 at.% lanthanide doped lead titanate samples from 100 to 550 °C (The curves are offset vertically for convenience).

The DSC results for PTLn8 samples are shown in figure 7, the peak associated with the FE-PE transition is again clearly observed. The depression of $T_c$ to 345 °C for PTLa8 is markedly larger than that observed for PTLa2. The transition temperature then recovers toward the UD value with decreasing ion size, reaching 445 °C for PTDy8. The recovery of $T_c$ is similar to the trend in tetragonality observed in figure 5.

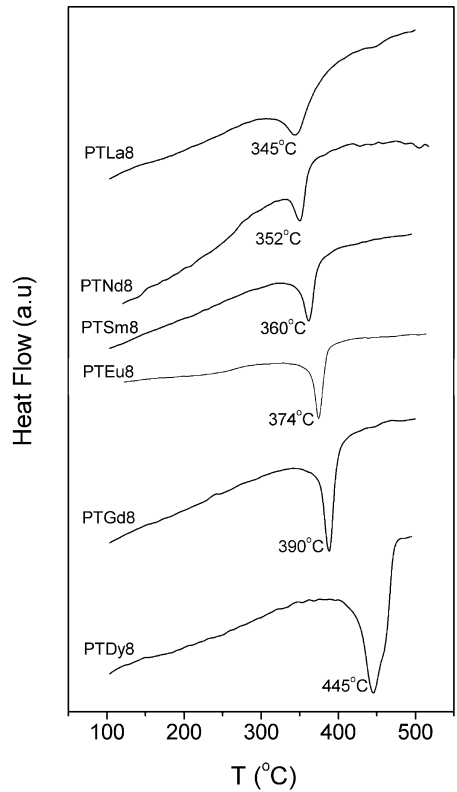

**Figure 7.** The differential scanning calorimetry curves for the 8 at.% lanthanide doped lead titanate samples from 100 to 550 °C (The curves are offset vertically for convenience).

## 4. Conclusions

The incorporation of lanthanide ions into lead titanate was studied using starting compositions that assumed donor doping, with the ions lanthanum, neodymium, samarium, europium, gadolinium and dysprosium, at 2 and 8 at.%. A suppression of the tetragonality and cell volume, and a concomitant decrease in the ferroelectric to paraelectric transition temperature, compared to undoped lead titanate was observed with La doping, consistent with substitution of $La^{3+}$ ions at the A-site. The cell volume was observed to reach a minimum for Nd incorporation, and then recover toward the undoped value with decreasing ion size. For the 2 at.% doped samples the tetragonality recover strongly for Gd and Dy, while for the 8 at.% samples the recovery began at Nd. In both cases the trend in the recovery of tetragonality was similar to that observed for the PE-FE transition temperature. The observations are consistent with the onset of fractional occupancy of the lanthanide ion on the B-site with decreasing ion size.

**Acknowledgements**

AP-B and YM-G wish to thank the Third World Academy of Sciences (RG/PHYS/LA Nos. 99-050, 02-225 and 05-043), to the ICTP, Trieste-Italy, for financial support of Latin-American Network of Ferroelectric Materials (NET-43), and to thank to R. de Lahaye Torres for sample preparation. AP-B and DJK thank the Royal Society of London for short-term visitor and international travel awards. AP-B acknowledges to the Conseil Régional Languedoc-Roussillon for her invitation in the University of Nîmes, France.